\documentclass[twocolumn,showpacs,amsmath,amssymb,prb,%
               superscriptaddress
              ]{revtex4}
\usepackage[T1]{fontenc}
\usepackage{times}
\usepackage{graphicx}
\usepackage{wasysym}
\usepackage{bm}% bold math

\begin{document}

\title{Isotropic three-dimensional left-handed meta-materials}

\author{Th.~Koschny}
\affiliation{Ames Laboratory and Dept.~of Phys.~and Astronomy,
             Iowa State University, Ames, Iowa 50011, U.S.A.}
\affiliation{Institute of Electronic Structure and Laser, FORTH,
             71110 Heraklion, Crete, Greece}

\author{L.~Zhang}
\affiliation{Ames Laboratory and Dept.~of Phys.~and Astronomy,
             Iowa State University, Ames, Iowa 50011, U.S.A.}

\author{C.~M.~Soukoulis}
\affiliation{Ames Laboratory and Dept.~of Phys.~and Astronomy,
             Iowa State University, Ames, Iowa 50011, U.S.A.}
\affiliation{Institute of Electronic Structure and Laser, FORTH,
             71110 Heraklion, Crete, Greece}

\date{\today}

\begin{abstract}
We investigate three-dimensional left-handed and related meta-materials based 
on a fully symmetric multi-gap single-ring SRR design and crossing continuous 
wires. We demonstrate isotropic transmission properties of a SRR-only 
meta-material and the corresponding left-handed material which possesses 
a negative effective index of refraction due to simultaneously negative
effective permeability and permittivity.
Minor deviations from complete isotropy are due to the finite thickness
of the meta-material.
\end{abstract}

% 41.20.Jb  Electromagnetic wave propagation; radiowave propagation 
% 42.25.Bs  Wave propagation, transmission and absorption 
% 42.70.Qs  Photonic bandgap materials 
% 73.20.Mf  Collective excitations


\maketitle

The realization of a perfect lens \cite{Pendry00} and other applications
of negative refraction requires the fabrication
of three-dimensional, homogeneous, isotropic left-handed 
materials\cite{Veselago68} (LHM) with simultaneously negative permittivity 
$\varepsilon$ and magnetic permeability $\mu$.
So far, no such materials exist, neither in nature nor in the laboratory.
Today's available LHM structures, based on the periodic arrangement of 
split ring resonators\cite{Pendry99} (SRR) and continuous metallic 
wires\cite{Pendry96b}, are only one 
dimensional\cite{Smith00,Parazzoli03,Aydin04} (1d),
supporting left-handed properties only for propagation with fixed 
polarization in one direction, 
or two dimensional\cite{Shelby01a,Shelby01b,Parazzoli04} 
(2d), where propagation in two directions with fixed polarization or 
one direction with arbitrary polarization is possible.
Earlier attempts to designs at least isotropic SRR\cite{GayBalmaz02a}
where lacking the symmetry of SRR and unit cell and required individual
tuning of the parameters in the different spatial directions.

In this paper, we propose a three dimensional (3d) isotropic LHM design 
which allows left-handed behavior for any direction of propagation and 
any polarization of the electromagnetic wave.
Using numerical transfer matrix simulations, 
we verify the isotropic transmission properties of the proposed structures.
Our data shows excellent agreement with results expected for a 
homogeneous slab with the corresponding negative $\varepsilon$ and $\mu$.

\begin{figure}
%\centerline{\includegraphics[width=8.5cm]{build/geometry/wrapper.eps}}
 \centerline{\includegraphics[width=8.5cm]{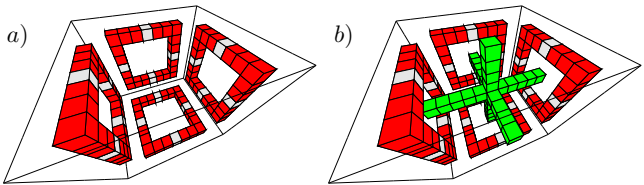}}
 \caption{%
 (Color online)
 The design of a fully symmetric unit cell for an one unit cell thick slab
 of an isotropic SRR ($a$) and a left-handed ($b$) meta-material.
 The interfaces are parallel to the left and right SRR.
% , periodic boundary
% conditions apply in the other two directions.
% The additional SRR in the second interface is required to preserve the 
% inversion symmetry of the meta-material in the direction
%  normal to the interfaces.
 The metal of the 4-gap SRR (red/dark gray) and the continuous wires 
 (green/medium gray) is 
 silver using a Drude-model permittivity around 1 THz.
 The SRR-gaps are filled with a high-constant dielectric (light gray) 
 with a relative permittivity $\varepsilon_\mathrm{gap}=300$
 to lower the magnetic resonance frequency. 
 }
 \label{fig:geometry}
\end{figure}

\begin{figure}
%\centerline{\includegraphics[width=8.5cm]{build/inversion/wrapper.eps}}
 \centerline{\includegraphics[width=8.5cm]{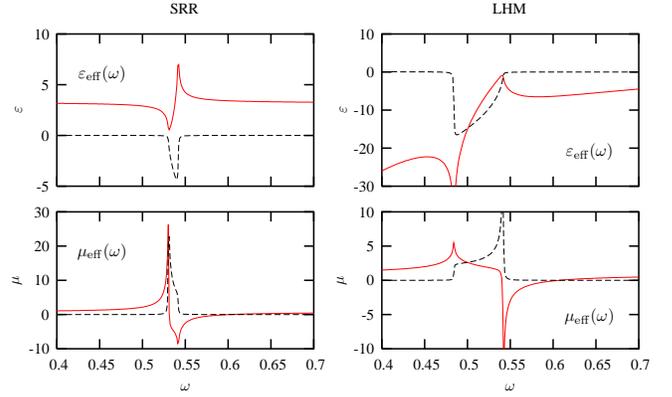}}
 \caption{%
 (Color online)
 The retrieved real (solid line) and imaginary (dashed line) part of the
 effective $\varepsilon_\mathrm{eff}(\omega)$
 and $\mu_\mathrm{eff}(\omega)$ of the homogeneous 
 medium approximation for normal incidence ($\vartheta=0$) to a single 
 unit cell layer of the SRR and LHM meta-material as a function of
 the dimensionless frequency $\omega=2\pi L f/c$. 
 }
 \label{fig:inversion}
\end{figure}

Our meta-materials are defined as a 3d periodic continuation 
of a single rectangular unit cell, consisting of SRRs and continuous wires. 
The sample is a slab of meta-material with a finite 
thickness of an integral number of unit cells and infinite extent in the 
perpendicular directions. The two surfaces of the slab are parallel to any 
face of the unit cell. 
An incident electromagnetic plane wave with wave vector $\bf k$ 
can be characterized by two angles:
the incidence angle $\vartheta\in[0,\pi/2)$ between $\bf k$ and the surface 
normal $\bf n$ of the sample, 
and the angle $\phi\in(-\pi,\pi]$ between the projection of $\bf k$ 
into and some chosen edge of the unit cell inside the surface plane 
of the sample.   
The frequency of the incident wave is chosen such that the vacuum wavelength
is approximately 10 times larger than the linear size of the unit cell
and we expect effective medium behavior.

To achieve an isotropic meta-material it is wise to start with 
a cubic unit cell.
The SRR and the continuous wire provide a resonant negative $\mu$ 
and a negative plasmonic $\varepsilon$, respectively. 
This occurs only when there is a component of the magnetic field 
normal to the SRR plane and a component of the electric field 
parallel to the wire. 
Therefore we have to use one pair of SRR and wire for 
each of the three spatial directions. 
The idea is that if the interaction between the SRR and wire in one direction,
and the mutual interaction in between the constituents in the different 
directions is negligible, we can, in first order, control the three diagonal 
elements of the permittivity and permeability tensors 
independently.
Previous results and extensive numerical simulations have further 
provided a set of design criteria to attain the best possible isotropy:

\smallskip\noindent
$(i)~$
To avoid the coupling of the electric field to the magnetic 
resonance\cite{Katsarakis04a,Marques02} of the circular current in the SRR 
we have to provide mirror symmetry of the SRR plain with respect to the 
direction of the electric field. 
Otherwise the electric resonant response of the magnetic resonance or the
emerging cross-polarization terms in more-dimensional media may destroy
the desired left-handed behavior.  
The problem can be avoided in 1d and certain 2d
(1d propagation with arbitrary polarization, 
 2d propagation with fixed polarization) meta-materials. 
However, for an isotropic material with arbitrary direction and 
polarization of the incident wave this can only be guaranteed by an 
inversion symmetric SRR design. 

\smallskip\noindent
$(ii)~$
From an isotropic, homogeneous effective medium we do not expect any 
cross-polarization scattering amplitudes.
Suppression of the cross-polarization terms requires, besides an isotropic
material discretization\cite{Koschny04e},
that the inversion symmetry of the unit cell is not simultaneously broken 
in both directions perpendicular to the direction of propagation. 
Therefore we have to center the SRRs on the faces of the unit cell and the
wires with respect to the positions of the SRRs.

\smallskip\noindent
$(iii)~$
For the finite slab a description in terms of an homogeneous effective 
medium is only possible if the isotropy, ie.~the inversion symmetry 
of the slab, in the direction of the finite dimension is 
preserved\cite{Koschny04c,Smith04b}.
As a consequence, we have to terminate the second surface of the slab by a 
repetition of the opposite surface.  
This is illustrated in Fig.~\ref{fig:geometry} 
for a sample of one unit cell thickness.

\smallskip\noindent
$(iv)~$
To avoid effects of the periodicity it is imperative to have the magnetic 
resonance frequency well below the first periodicity band-gap. 
But even in this case we observe artifacts\cite{Koschny03b} in the 
effective medium response, like resonance/anti-resonance coupling and negative
imaginary parts.
These issues have been addressed in detail in Ref.~\onlinecite{Koschny04c}.
Further we have to make sure that the magnetic resonance frequency is 
below the effective plasma frequency in the LHM, which is reduced compared
to the plasma frequency of the wires by the additional electric (cut-wire)
response of the SRR\cite{Koschny04a,Katsarakis04b}.

\smallskip\noindent
$(v)~$
In the LHM, 
%the above demands for inversion symmetry leave us basically two
%possibilities to place the continuous wires: 
%along the edges of the unit cell or intersecting in the middle of the 
%unit cell as shown in Fig.~\ref{fig:geometry}.
the best position for the wires has been found to be aligned with the middle 
of the SRR in the center between a SRR and its periodic continuation. 
This design minimizes the disturbance between SRRs and wires.
Additionally, it is favorable to put the edges of the SRRs not too close 
to one another to avoid capacitive coupling across adjacent gaps.

\smallskip
To meet the symmetry requirements of the SRR we deploy a symmetric 
single-ring design with the gap distributed over all four sides as
shown if Fig.~\ref{fig:geometry}.
The 4-gap single-ring SRR has been proven as a simpler, symmetric and
working alternative to the conventional design using nested rings with
a single gap each\cite{OBrien04}.
The gap-width in the 4-gap SRR has to be significantly reduced in comparison 
to a 1-gap design since due to the receded capacitance of the serial 
capacitors the resonance frequency of the SRR would increase.
Alternatively, we could increase the dielectric constant inside the gaps
to move the resonance frequency down.
In the simulations presented in this paper we chose the latter strategy.

\begin{figure}
%\centerline{\includegraphics[width=8.5cm]{build/1uc-sym/wrapper.eps}}
 \centerline{\includegraphics[width=8.5cm]{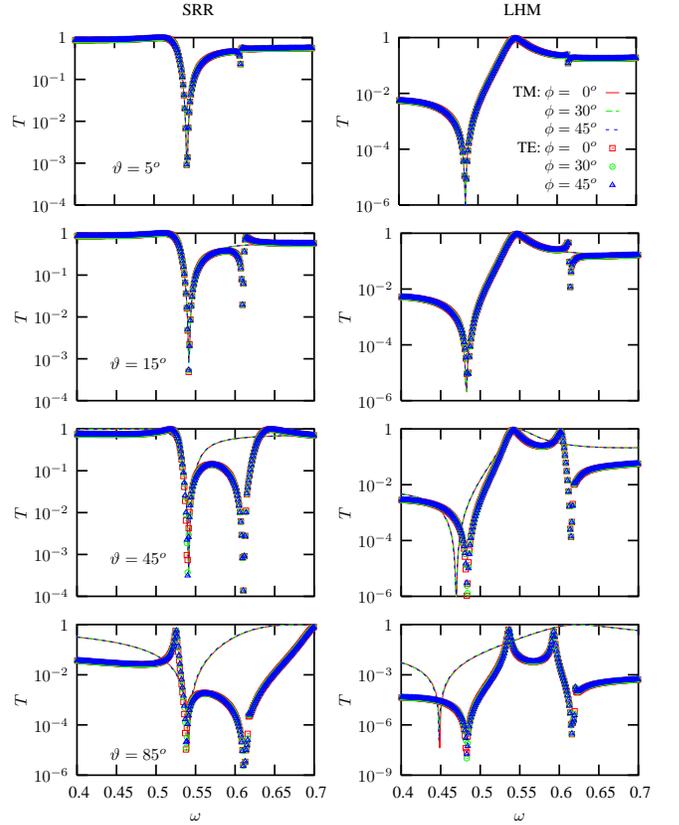}}
 \caption{%
 (Color online)
 Transmission spectra $T=|t(\omega)|^2$ for a single layer of unit cells
 of the SRR and the LHM meta-material for various angles of
 incidence ($\vartheta$) and polarizations ($\phi$, TE and TM mode).
 }
 \label{fig:tr1uc}
\end{figure}

\begin{figure}
%\centerline{\includegraphics[width=8.5cm]{build/ana-TR/wrapper-1uc.eps}}
 \centerline{\includegraphics[width=8.5cm]{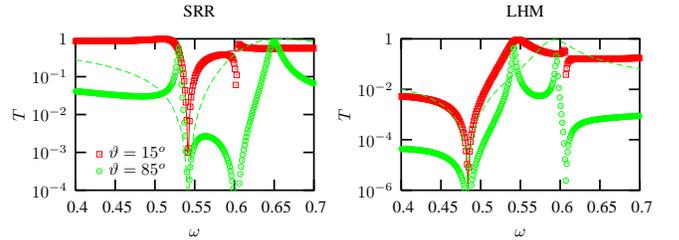}}
 \caption{%
 (Color online)
 Analytic transmission spectra $T=|t(\omega)|^2$ of the TE and TM mode
 for a single layer of {\em homogeneous} unit cells of the SRR and LHM
 material for two angles of incidence.
 For $\mu(\omega)$ and $\varepsilon(\omega)$ we used the retrieved
 values for normal incidence (Fig.~\ref{fig:inversion}).
 }
 \label{fig:tr1uc_ana}
\end{figure}

\begin{figure}
%\centerline{\includegraphics[width=8.5cm]{build/muc-simple/wrapper.eps}}
 \centerline{\includegraphics[width=8.5cm]{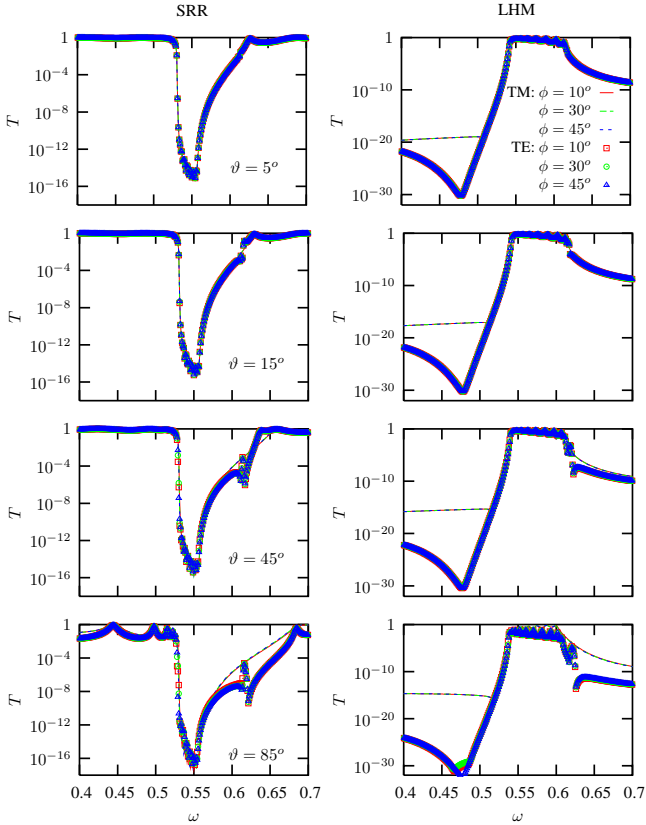}}
 \caption{%
 (Color online)
 Transmission spectra $T=|t(\omega)|^2$ for a 10 unit cell thick slab
 of the SRR and the LHM meta-material for various angles of
 incidence ($\vartheta$) and polarizations ($\phi$, TE and TM mode).
 }
 \label{fig:tr10uc}
\end{figure}

\begin{figure}
%\centerline{\includegraphics[width=8.5cm]{build/ana-TR/wrapper-10uc.eps}}
 \centerline{\includegraphics[width=8.5cm]{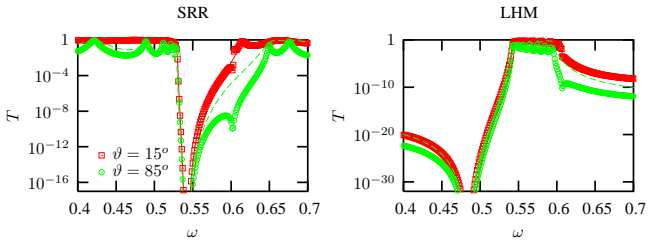}}
 \caption{%
 (Color online)
 Same as Fig.~\ref{fig:tr1uc_ana} but for 10 unit cell thickness.
 }
 \label{fig:tr10uc_ana}
\end{figure}

For the numerical simulations we employ a lattice transfermatrix technique 
as described in \cite{Pendry96,Markos02a} to calculate the complex scattering 
amplitudes of the sample. 
The finite discretization in conjunction with periodic boundary conditions
does only allow a finite set of values for the component of the wavevector 
parallel to the surface, hence only a finite set of incidence directions. 
As a consequence, 
to calculated the scattering amplitudes for an incident 
plane wave under an arbitrary incidence angle $\vartheta$ we have 
to impose more general, quasi-periodic boundary conditions to the unit cell 
in both directions perpendicular to the stratification direction.
This means the electromagnetic fields at either side of the unit cell 
are identified up to a phase $e^{i Q_m L}$ where $m$ enumerates the
perpendicular directions, $Q_m\in[0,2\pi/L)$ are the Bloch momenta parallel 
to the surface, and $L$ is the size of the unit cell.
Since the parallel momentum is preserved across the surfaces, 
we can calculate the $Q_m$ from the angles of incidence, 
$Q_1=(k_\|\,\mathrm{mod}\, 2\pi/L) \cos\phi$, 
$Q_2=(k_\|\,\mathrm{mod}\, 2\pi/L) \sin\phi$.

We calculate the transmission and reflection
amplitudes for a slab of the isotropic SRR and the corresponding isotropic 
LHM meta-material of one and ten unit cells thickness.
The metals are implemented using a Drude-model permittivity for silver
around 1 THz. The unit cell size is chosen accordingly to facilitate a 
resonance frequency in the same region.
All results are given in terms of the dimensionless frequency 
$\omega=2\pi fL/c$ since up to the low terahertz region the scaling of the 
meta-materials resonance frequencies is linear and virtually independent 
on the permittivity of the metals. 

Figure \ref{fig:inversion} shows the retrieved $\varepsilon(\omega)$
and $\mu(\omega)$ of the effective homogeneous medium approximation of the
SRR and LHM slab of one unit cell thickness for normal incidence, 
proving the presence of the magnetic resonance featuring $\mu<0$ in the SRR
and the left-handed bandpass with $\varepsilon<0, \mu<0$ in the LHM.
The retrieved $\varepsilon(\omega)$ and $\mu(\omega)$ 
% for normal incidence 
are, in good approximation, independent on the thickness of the slab.
The clearly visible resonance/anti-resonance coupling and the occurrence 
of negative imaginary parts are artifacts arising from the 
periodicity\cite{Koschny04c} which could be reduced by moving the magnetic 
resonance to sufficiently low frequency.

Figures \ref{fig:tr1uc} and \ref{fig:tr10uc} show the simulated transmission 
$|t(\omega)|^2$ around the magnetic resonance for the one and ten unit cell 
thick slab, respectively, for different incident angles $\vartheta$ and 
polarizations. Because of the unit cell's symmetry it is sufficient to
consider $\phi\in[0,\pi/4]$ for the TE and TM mode.
Despite the square shape of the SRR, we find absolute independence of the
scattering amplitudes (reflection not shown) on the orientation $\phi$ of the
incidence plane for both, TE and TM mode, arbitrary $\vartheta$, SRR and LHM
meta-material slabs of any thickness.
This finding is important for experimental realizations as it demonstrates
that a circular shape of the SRRs is not required for isotropic results.
As expected for an homogeneous slab, the transmission spectra depend on the 
incidence angle $\vartheta$, and, for non-normal incidence, also differ
for the TE and TM polarization. 
In accordance with the retrieved $\varepsilon$ and $\mu$, the SRR spectra 
show a transmission dip at the magnetic resonance frequency $\approx 0.54$,
which widens into a stop-band for the longer slab, 
while the LHM spectra show the corresponding transmission peak, developing
into a well-defined LH passband for the longer system.
This behavior prevails from normal to nearly parallel incidence.
Surprisingly, for off-normal incidence a second spectral feature emerges
for both, SRR and LHM meta-material at around $\omega\approx 0.61$ but only
in the TE polarization, which gets less accentuated for the longer slab.
However, this turns out to be an expected feature for an isotropic, homogeneous
slab and has a simple explanation:
The analytic transmission amplitude has the form 
$t^{-1}=\cos qd-(i/2)(\zeta+1/\zeta)\sin qd$, where $q$ is the component of
the wavevector perpendicular to the surface inside the slab,
$k$ the corresponding component in vacuum, and $\zeta=\mu k/q$ for the TE 
and $\zeta=\varepsilon k/q$ for the TM mode, what becomes the impedance of the
slab for the case of normal incidence.
% Since the parallel momentum in preserved, 
$q$ is given by the dispersion 
relation $q^2+k_\|^2-\mu\varepsilon\,\omega^2/c^2=0$.
Whenever $\mu(\omega)$ for the TE mode or $\varepsilon(\omega)$ for the TM
mode becomes zero, the prefactor of the sine in $t^{-1}$ diverges. 
For normal incidence, ie.\@ $k_\|=0$, this divergence is compensated by the 
vanishing $q\propto(\mu\varepsilon)^{1/2}\to 0$.
For $\vartheta>0$, however, $k_\|\not=0$ renders the sine non-zero such that
the divergence of the prefactor leads to a dip in the transmission $t$.
The observed dip for the TE mode in the simulations indeed clearly corresponds
to the zero of the permeability present in Fig.~\ref{fig:inversion}, further 
supporting the isotropic behavior of the meta-material slabs.

Figures \ref{fig:tr1uc_ana} and \ref{fig:tr10uc_ana} show the transmission 
spectra analytically calculated for homogeneous slabs of the corresponding
lengths, assuming homogeneous $\varepsilon(\omega)$, $\mu(\omega)$ taken 
from the results shown in Fig.~\ref{fig:inversion}.
The similarity to the simulation results, particularly for the SRR case, is
striking, again confirming the isotropic behavior of the 
meta-material slabs.
Slight deviation of the simulated transmission for both, SRR and LHM, 
can be attributed to two principal deficiencies of our estimation of
$\varepsilon(\omega)$ and $\mu(\omega)$: 
First, the artifacts arising from the periodicity will introduce an explicit
dependence of the incident angle $\vartheta$ since the unit cell is rectangular
(and we expect the band structure to look different for various 
directions). Second, due to the symmetric surface termination the 
relevant periodicity for the retrieval shown in Fig.~\ref{fig:inversion} is
one mesh step longer than the period in the perpendicular directions.
These effects will become negligible if we move the magnetic resonance 
to sufficiently low frequency while maintaining the size of the unit cell.

The LHM transmission spectra reveal an additional, more serious problem. 
The finite thickness of the slab leads to an essential asymmetry in the 
electric response between the directions parallel and normal
to the surface. The electric field in the normal direction sees a finite, 
instead of a continuous wire, which changes the electric response, hence 
$\varepsilon(\omega)$ in that direction, 
leading to an deformed spectrum for large $\vartheta$.
This problem only affects the TM mode for oblique incidence.
As is apparent from Fig.~\ref{fig:tr10uc}, within the LH passband 
we can expect this issue to become less important with increasing 
thickness of the slab. 
However, in the transmission gaps the near unity reflection is dominated 
by the few surface layers of the slab, thus experiencing the termination of 
the finite wires for any system length.

We are aware that the geometry of the SRR with dielectric-filled gaps 
shown in Fig.~\ref{fig:geometry} is not very convenient for experimental 
realizations.
However, the issues discussed in this paper are generic and do not require
this particular realization of the SRRs.
Experimentally more realistic SRRs that can operate at very low frequency
not sacrificing the symmetry can be designed based on multi-layer 4-gap 
structures or two-layer opposing conventional SRRs\cite{Marques02}.
For low frequencies also the wires may be thinner, such that the crossing can 
easily replaced by centered wires which just touch each other in the center
of the unit cell.

In conclusion, we presented a design for an LHM and SRR-only meta-material 
based on a fully symmetric unit cell.
Transfermatrix simulations of finite thickness slabs
% of these meta-materials 
for oblique incident plain waves revealed almost isotropic transmission 
spectra. Further, the simultaneously negative $\varepsilon$ and $\mu$
has been confirmed for the LHM slab.
We emphasized the importance of symmetry issues in the design of isotropic
meta-materials. Especially the asymmetry introduced by the gaps in the 
conventional SRR design has to be avoided.
Additionally, the magnetic resonance frequency has to be chosen much lower
than the first band gap arising from the meta-materials intrinsic periodicity
to avoid interference with the band structure and the breakdown of the 
effective medium behavior.
Minor deviations from the isotropic behavior which occur particularly in 
the LHM slab have been attributed to the finite thickness of the meta-material
which modifies the electric response in the short direction.
Most of the imperfections become unimportant for sufficiently long systems.

This work was partially supported by Ames Laboratory 
(Contract number W-7405-Eng-82). 
Financial support of EU$\underline{~~}$FET project DALHM 
and DARPA (Contract number  MDA972-01-2-0016)
are also acknowledged.
The authors thank Peter Marko\v{s} for valuable discussions.

\bibliographystyle{apsrev}

\end{document}